\documentclass[%        	Class options:
notitlepage,
letterpaper,
preprintnumbers,
amsmath,
amssymb,
superscriptaddress,%    	Authors' addresses linked with superscripts
nofootinbib,%           	Does not treat footnotes as references
floatfix,%                	Fixes float errors
hyperref]%						hyper references
{revtex4-1}%              	REVTEX 4 Package used

\usepackage{graphicx,%  	Default Latex 2eps package for embedding figures
longtable}%             	Useful for long table
\usepackage{amsmath,amstext,amsfonts,amsbsy,amssymb,amscd,bbm,epsfig}
\usepackage[breaklinks,colorlinks,urlcolor=blue,citecolor=blue,linkcolor=blue]{hyperref}
\usepackage{natbib}
\usepackage{url}

\usepackage{color}
\usepackage{bm}
\usepackage{rotating}
\usepackage{latexsym}
\usepackage{multirow}

\begin{document}

\title{Neutrino tomography of the Earth}

\preprint{\tt IFIC/17-52}

\author{A.~Donini$^1$, S. Palomares-Ruiz$^1$ and J. Salvado}
\affiliation{
Instituto de F\'{\i}sica Corpuscular, CSIC-Universitat de Val\`encia,\\
       Apartado de Correos 22085, E-46071 Val\`encia, Spain }
\affiliation{ Institut de Ci\`encies del Cosmos, Universitat de Barcelona, \\
 Diagonal 647, E-08028 Barcelona, Spain}

\begin{abstract}
Cosmic-ray interactions with the nuclei of the Earth's atmosphere produce a flux of neutrinos in all directions with energies extending above the TeV scale~\cite{Gaisser:2002jj}. However, the Earth is not a fully transparent medium for neutrinos with energies above a few TeV. At these energies, the charged-current neutrino-nucleon cross section is large enough so that the neutrino mean-free path in a medium with the Earth's density is comparable to the Earth's diameter~\cite{Gandhi:1995tf}. Therefore, when neutrinos of these energies cross the Earth, there is a non-negligible probability for them to be absorbed. Since this effect depends on the distance traveled by neutrinos and on their energy, studying the zenith and energy distributions of TeV atmospheric neutrinos passing through the Earth offers an opportunity to infer the Earth's density profile~\cite{GonzalezGarcia:2007gg, Borriello:2009ad, Takeuchi:2010, Romero:2011zzb}. Here we perform an Earth tomography with neutrinos using actual data, the publicly available one-year through-going muon sample of the atmospheric neutrino data of the IceCube neutrino telescope~\cite{TheIceCube:2016oqi}. We are able to determine the mass of the Earth, its moment of inertia, the mass of the Earth's core and to establish the core is denser than the mantle, using weak interactions only, in a way completely independent from gravitational measurements. Our results confirm that this can be achieved with current neutrino detectors. This method to study the Earth's internal structure, complementary to the traditional one from geophysics based on seismological data, is starting to provide useful information and it could become competitive as soon as more statistics is available thanks to the current and larger future neutrino detectors.
\end{abstract}

\maketitle

\section{Introduction}

A reliable estimate of the density profile of the Earth is essential to solve a number of important problems in geophysics, such as the dynamics of the core and mantle, the mechanism of the geomagnetic dynamo or the bulk composition of the Earth~\cite{Bolt:1991}. Most of our knowledge about the internal structure of the Earth and the physical properties of its different layers comes from geophysics and, in particular, from seismological data. Moreover, information from geomagnetic and geodynamical data, solid state theory and high temperature/pressure experimental results is also used. 

The determination of the density distribution of the Earth from bulk sound velocity of seismic waves in combination with normal modes is a well-established method with statistical uncertainties in the mantle at the few percent level and larger errors for core densities~\cite{Kennett:1998, Masters:2003} . The reconstruction of a three-dimensional profile is, however, a very demanding non-linear inversion problem of different seismic data~\cite{Kennett:1998, Masters:2003, deWit:2014}. Moreover, as wave velocities also depend on composition, temperature, pressure and elastic properties, this necessarily introduces uncertainties in the density estimate. Most studies of the Earth'€™s radial structure are based on empirical relations between seismic waves velocities and density such as the Birch's law, which may fail at the higher densities of the Earth's core, and the Adams-Williamson equation~\cite{AWequation}. A good understanding of the Earth's  interior, aiming at simultaneously determining the density variations and the origin of such waves in terms of temperature and composition variations, cannot be done from seismic velocities variations alone and another, independent piece of information is needed. Therefore, a precise modeling of the different layers composition which are crossed by seismic waves is required. Even though several million of earthquakes occur in the Earth every year, only of the order of hundred of them have magnitudes larger than 6~\cite{earthquakes}. Most of them do not occur on the surface, and the origin of the wave must be inferred by comparing time delays from different seismographs. Eventually, only a small fraction of the registered seismic waves cross the Earth's core. For all these reasons, using other complementary and independent methods to infer the density profile of the Earth is important.

\begin{figure*}
	%	\begin{center}
	\begin{tabular}{cc}
		\includegraphics[width=0.4\textwidth]{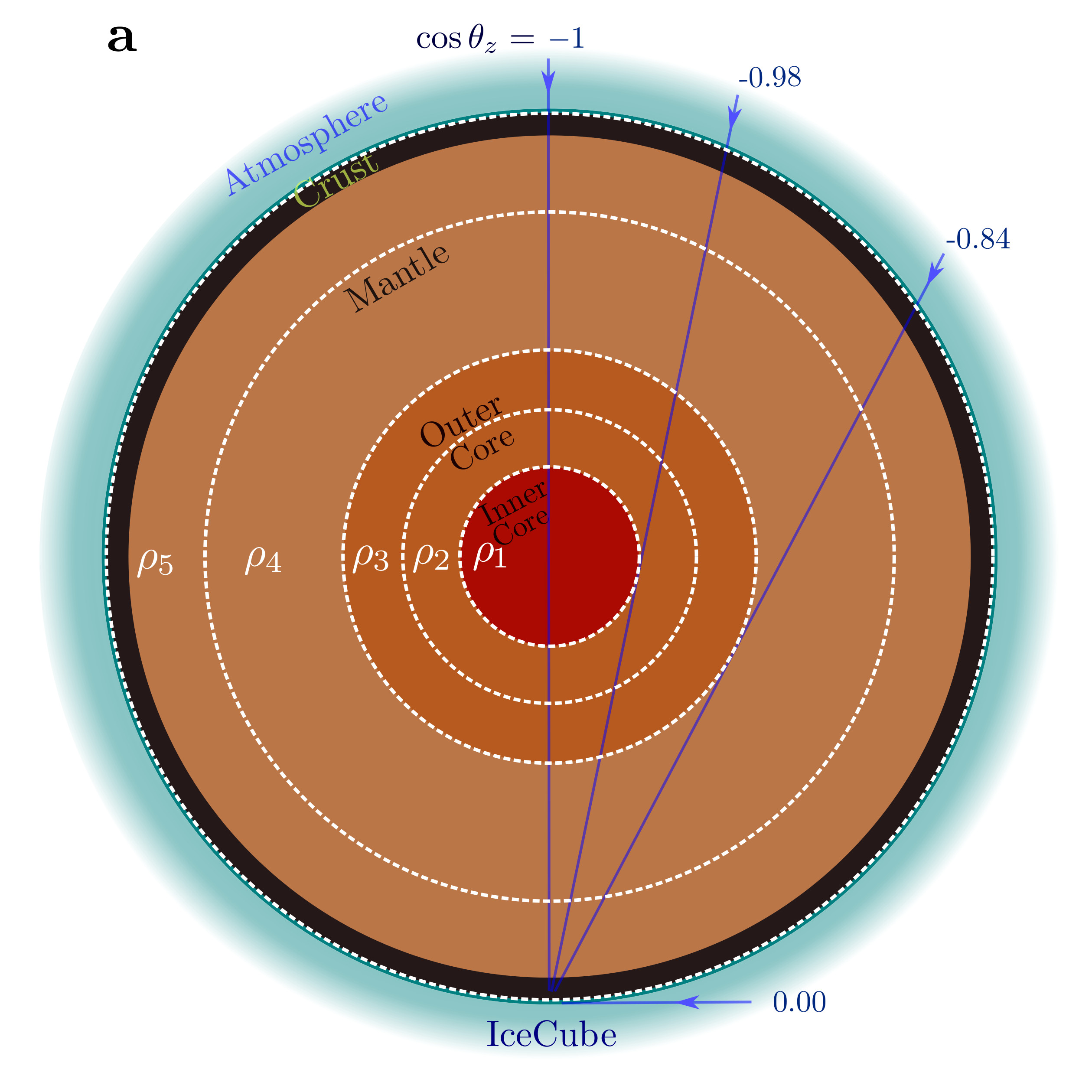} & \hspace{1cm}
		\includegraphics[width=0.38\textwidth]{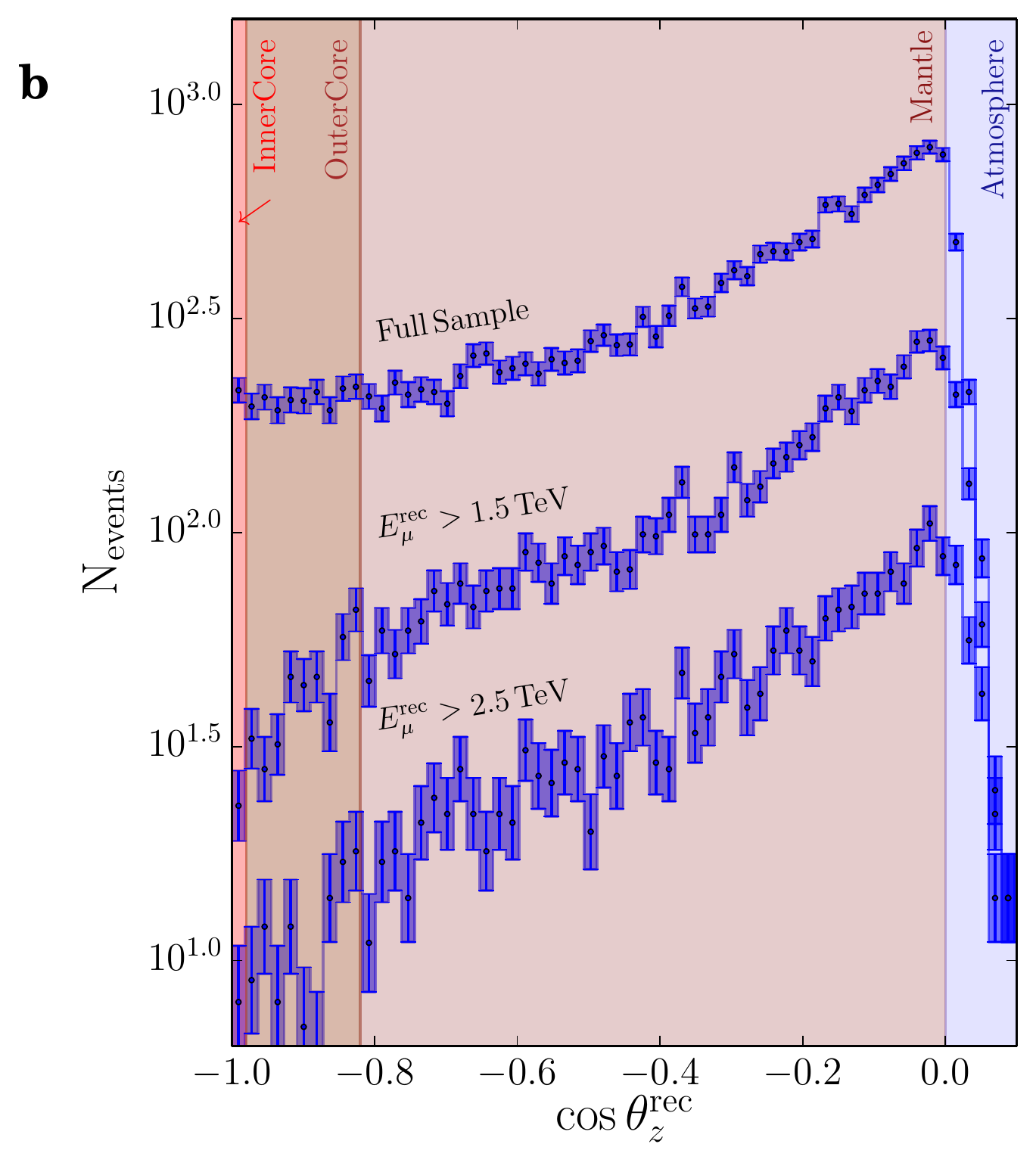} 
	\end{tabular}
	\caption{
		{\bf Zenith angular distribution of the atmospheric muon neutrino events in the IC86 sample.} 
		{\bf a,} Schematic pictorial representation of the Earth subdivided in the five concentric layers used in this work. Some representative neutrino trajectories and their associated zenith angles, $\theta_z$, with respect to the IceCube detector are also indicated.
		{\bf b,} Number of atmospheric up-going muon neutrino events collected in the first year of IceCube data taking as a function of the cosine of the zenith angle $\theta_z^{\rm rec}$ for different reconstructed muon energy thresholds. The uppermost curve shows the zenith distribution for the entire IC86 sample (i.e., 20145 muons in the energy range $400~{\rm GeV} < E_\mu^{\rm rec} < 20$~TeV) and the lowermost curve corresponds to the highest threshold in this plot, $E_\mu^{\rm rec} > 2.5$~TeV. Up-going neutrinos correspond to $\cos \theta_z^{\rm rec} = -1$.
	}
	\label{fig:zenith}
	%	\end{center}
\end{figure*}

Neutrinos can be used to study the Earth's interior in several ways. First of all, experiments such as KamLAND and Borexino are currently measuring the so-called geo-neutrino flux (i.e., neutrinos produced by the decay of radioactive elements in the Earth's interior~\cite{Gando:1900zz, Bellini:2010hy}), which provides information that can be used to understand its composition. On the other hand, a good knowledge of neutrino propagation through the Earth may give relevant information about the Earth's density profile. Neutrino propagation does depend, indeed, on the details of the matter structure between the source and the detector. For neutrinos with energy below 1~TeV, the matter profile affects the neutrino oscillation pattern~\cite{Rott:2015kwa, Winter:2015zwx, Bourret:2017tkw}, whereas for neutrinos with energies in the multi-TeV range, the neutrino flux observed at the detector depends on the number of nucleons along its path, as neutrinos can undergo inelastic scattering and get absorbed. Indeed, the idea of performing absorption radiographies of the Earth with neutrinos dates back to more than four decades ago. To our knowledge, the first mention of this possibility was advanced in an unpublished CERN preprint in October 1973 by Placci and Zavattini~\cite{Placci:1973} and by Volkova and Zatsepin in a talk of 1974~\cite{Volkova:1974xa}, considering man-made neutrinos. The idea of combining neutrino Earth's radiographies, i.e., performing a neutrino tomography, is based on the study of the attenuation of neutrinos crossing the Earth from different angles with respect to the position of the detector. The column depth traversed by a neutrino that has passed through the entire Earth's diameter is 11 kton/cm$^2$ ($1.1 \times 10^{10}$~cmwe). For neutrinos with an energy of $\sim 40$~TeV, the absorption length in the Earth becomes comparable to its diameter, $\left(n \, \sigma\right)^{-1} \sim 2 \, R_\oplus$, where $n$ is the average nucleon number density, $\sigma$ the neutrino-nucleon total cross section and $R_\oplus = 6371$~km the mean radius of the Earth. Therefore, for few TeV neutrinos there is a non-negligible probability for the incoming neutrino flux to be suppressed, $e^{-n \, \sigma \, L} < 1$, where $L = 2 \, R_\oplus \, \cos \theta_z$ is the path length in the Earth as a function of the zenith angle $\theta_z$ (Fig.~\ref{fig:zenith}a).

%%%%%%%%%%%%
%%%%%%%%%%%%

\section{Data and methods}

\subsection{IC86 atmospheric neutrino sample} 

Atmospheric neutrinos offer a large range of baselines (from a few to thousands of kilometers) and energies (from MeV to tens of TeV), with an energy spectrum that falls as $\sim E^{-3.7}$.  Therefore, they represent a suitable source for neutrino tomography. Although neutrino interactions are rare, with the operation of kilometer-cube detectors such as IceCube, a large event sample can be harvested. In this work we use the publicly available IceCube one-year up-going muon sample,  collected during 2011-2012 and referred to as IC86 (IceCube 86-string configuration), which contains 20145 muons detected over a live time of 343.7 days~\cite{TheIceCube:2016oqi} (a preliminary attempt using IceCube data with very limited event statistics was presented in 2012~\cite{ICradiography}). These muons are produced by up-going neutrinos and antineutrinos which, after crossing the Earth, interact via charged-current processes in the bedrock or ice surrounding the detector. In turn, these neutrinos are originated from decays of atmospheric pions and kaons (and with a contamination from other sources below 0.1\%) in the Northern hemisphere, and have all traversed the Earth. The sample covers a solid angle of $2 \pi$, making it particularly suitable for the kind of study performed here. While propagating inside the detector at a speed higher than the speed of light in ice, these muons emit Cherenkov light, which is detected by the digital optical modules of the IceCube array. The energy of the muons in the IC86 sample lies between 400 GeV and 20~TeV and is reconstructed, based on energy losses along the track, with a resolution of $\sigma_{\rm \log (E_\mu/{\rm GeV})} \sim 0.5$. Since the median opening angle with respect to the parent neutrino direction is $0.7 \, (E_\nu/{\rm TeV})^{-0.7}$~degrees~\cite{TheIceCube:2016oqi}, the muon direction is a very good proxy for the original neutrino direction. The muon zenith angle can be reconstructed with a resolution in $\cos \theta_z$ between 0.005 and 0.015.

\subsection{Models of atmospheric neutrino fluxes}

The atmospheric neutrino flux is characterized in terms of the cosmic-ray primary spectrum entering the atmosphere and the hadronic interaction model that controls the development of the shower that finally produces the flux of neutrinos. Several choices for the model of the atmospheric neutrino flux are currently compatible with all available data: in this letter, we choose for our analysis results from the combined Honda-Gaisser model and Gaisser-Hillas H3a correction (HG-GH-H3a) for the primary cosmic-ray flux~\cite{Gaisser:2013bla} and the QGSJET-II-04 hadronic model~\cite{Ostapchenko:2010vb}. Nevertheless, we also considered the Zatsepin-Sokolskaya (ZS) cosmic-ray spectrum~\cite{Zatsepin:2006ci} and the SIBYILL2.3 hadronic model~\cite{Riehn:2015oba} and combined them obtaining a set of four different models for the atmospheric neutrino fluxes.

\subsection{Neutrino-nucleon cross sections}

In the energy range relevant for this analysis (i.e., neutrino energies between few hundred GeV and few tens of TeV), the neutrino-nucleon and antineutrino-nucleon cross sections are known\footnote{Note that taking a complementary approach to the one presented in this letter, one could try to confirm the value of the neutrino-nucleon cross section at these energies, assuming the Earth density profile to follow the PREM~\cite{Aartsen:2017kpd}.} within $(2-3)\%$ and $(4-10)\%$, respectively~\cite{CooperSarkar:2011pa}. In this work, we use the parton distribution functions from the HERAPDF set~\cite{Aaron:2009aa} as our default interaction model, but we have also checked that the effect of the uncertainties in the neutrino cross sections is subdominant over other sources of error, so we do not discuss them any further.

\subsection{Propagation of neutrinos through the Earth}

The transport equations for neutrinos traversing the Earth, which we solve using the $\nu$-SQuIDs package~\cite{Delgado:2014kpa, nusquids}, consist of four main ingredients (see, e.g., Ref.~\cite{GonzalezGarcia:2005xw}):
(1) the standard evolution Hamiltonian in matter, which includes the vacuum mass-mixing terms and the effect of coherent forward scattering off electrons of the medium, given by the matter interaction potential; 
(2) the attenuation effect caused by neutrino inelastic interactions with matter, either via charged-current or neutral-current processes; 
(3) the redistribution of neutrinos from higher to lower energies after neutral-current interactions, 
and, 
(4) the neutrino regeneration term from tau lepton decays. 
Neutrino flavor oscillations in matter, given by the first term, represent the dominant effect for neutrino energies below a few hundred GeV. On the other hand, the other terms become dominant for neutrinos with higher energies. Since the neutrino-nucleon cross section increases with energy, at these energies the neutrino flux gets attenuated~\cite{Gandhi:1995tf}. In the case of neutral-current interactions neutrinos are degraded in energy~\cite{Berezinsky:1986ij}. In the case of charged-current interactions neutrinos are absorbed and a lepton of the same flavor is produced. Whereas in the case of electron and muon neutrinos (and antineutrinos), the associated lepton (electron or muon) is rapidly absorbed in the Earth and does not contribute to the high-energy neutrino flux, the tau leptons produced after tau neutrino charged-current interactions decay before losing too much energy. In these decays, a new tau neutrino (or antineutrino) with lower energy is produced and thus, they get regenerated~\cite{Halzen:1998be}. Moreover, secondary electron and muon neutrinos (and antineutrinos) are also produced after tau lepton decays~\cite{Beacom:2001xn}. For the energies we consider here, neutrino oscillations are suppressed and the effects of tau neutrino regeneration and secondary production of electron and muon neutrinos are negligible. On one hand, tau neutrinos are rarely produced in the atmosphere and on the other hand, this effect is only important for spectra much harder than the atmospheric neutrino one. Therefore, for the sake of saving computational time, we have not included the regeneration or secondary production terms in this work. We stress the corrections are much smaller than the precision on the determination of the Earth's profile achieved with current data.

\subsection{Nuisance parameters}

To relate the true variables (muon energy and direction) to the reconstructed observables (deposited energy and track zenith angle) we use the high-statistics Monte Carlo released by the IceCube collaboration along with the data. This also allows us to do a realistic treatment of the detector systematic uncertainties. In order to do so, we consider four of the continuous nuisance parameters described in the IC86 paper~\cite{TheIceCube:2016oqi}, where we refer the reader for further details. 
(1) The overall flux normalization ($N$) is allowed to vary within a factor of 2 of the central value of each model, which is larger than current uncertainties~\cite{Barr:2006it, Fedynitch:2012fs}. The low-energy component of the observed neutrino spectrum is extremely effective to substantially reduce the normalization uncertainty, though.
(2) The pion-to-kaon ratio ($\pi/K$) determines the relative contribution to the neutrino flux from pion or kaon decays. A smaller value of this parameter implies a harder atmospheric neutrino spectrum. We normalize it to one and use a 10\% Gaussian prior.
(3) The uncertainty on the spectral shape of the atmospheric neutrino spectrum, $\Delta \gamma$, is accounted for by a tilt in the energy spectrum, with a pivot energy close to the median of the neutrino energy distribution. We add it as a Gaussian prior with a 5\% error.
(4) The uncertainty in the efficiency of the digital optical modules (DOM$_{\rm eff}$) affects the determination of the reconstructed energy: a smaller efficiency implies a shift to lower energies. We allow this parameter to vary freely between 0.9 and 1.19, given that its central value is 0.99.

\subsection{Earth modeling}

We have considered two models for the Earth, fixing the position of the core-mantle boundary and the transition from the inner to the outer core. On one hand, we have parametrized the Earth's density with a one-dimensional five-layer, $R_1, \dots, R_5$, with constant density in each of the layers (Fig.~\ref{fig:zenith}a). The layers are defined as follows: $R_1 \in [0,0.195]~R_\oplus$,  $R_2 \in [0.195, 0.3725]~R_\oplus$, $R_3 \in [0.3725, 0.55]~R_\oplus$, $R_4 \in [0.55, 0.775]~R_\oplus$, $R_5 \in [0.775, 1]~R_\oplus$, where $R_\oplus$ is the mean Earth's radius, $R_\oplus = 6371$ km. The first layer corresponds to the inner core, the second and third layers (of equal thickness) to the outer core and the last two layers (of equal thickness) to the mantle. The density of each of these layers is allowed to float freely and independently. On the other hand, we have also considered a model with five layers, again, but with a density profile within each layer that follows that of the PREM. The density in each of these layers is multiplied by a factor which is also allowed to vary freely and independently of the others. For the current data set, we do not expect that a larger number of layers would change the results presented here. Indeed, this is partly explained by the similarity of the results of the flat-layer model and the PREM-based model with five layers. Since our aim in this work is to evaluate the sensitivity of the neutrino attenuation effect to the Earth's density profile, throughout this work, we have not imposed any external constrain on the Earth's mass or moment of inertia, which are (gravitationally) known much more precisely than what currently can be achieved with neutrinos~\cite{Luzum:2011, AAlmanac, Chen:2015}.

\subsection{Parameter estimation}

To quantitatively assess the power of the one-year up-going muon IC86 sample to determine the Earth's density profile, we performed a likelihood analysis (using the MultiNest nested sampling algorithm~\cite{Feroz:2007kg, Feroz:2008xx, Feroz:2013hea}) using all the events in the data sample and characterizing each event by its reconstructed muon energy and zenith angle. The full likelihood is defined as the bin product of the Poisson probability of measuring $N^{\rm data}_i$ for the expected value $N^{\rm th}_i$ times the product of Gaussian probabilities for the pulls of the nuisance parameters. The log-likelihood (up to a constant) is given by
\begin{equation}
\label{eq:L}
\ln \mathcal{L} ({\boldsymbol \rho} \, ; {\boldsymbol \eta}) = \sum_{i\in {\rm bins}}  \left(N^{\rm data}_i \, \ln N^{\rm th}_i ({\boldsymbol \rho} \, ; {\boldsymbol \eta}) - N^{\rm th}_i ({\boldsymbol \rho} \, ; {\boldsymbol \eta})\right) 
- \sum_j \frac{(\eta_j-\eta^0_j)^2}{2 \, \sigma_j^2} ~,
\end{equation}
where the subindex $i$ refers to a bin in reconstructed muon energy ($E^{\rm rec}_\mu$) and cosine of the reconstructed zenith angle ($\cos \theta^{\rm rec}_z$); $N^{\rm th}_i ({\boldsymbol \rho} \, ; {\boldsymbol \eta})$ is the expected number of evens for a given value of the densities in each layer (parameterized by ${\boldsymbol \rho}$) and the nuisance (${\boldsymbol \eta} \equiv \{N, \pi/K, \Delta \gamma, {\rm DOM}_{\rm eff}\}$) parameters in the $i$-th bin; and $N^{\rm data}_i$ is the number of data events in the same $i$-th bin. The index $j$ corresponds to the nuisance parameters with Gaussian prior ($\pi/K$ and $\Delta \gamma$) and $\sigma_j$ is the Gaussian error. To compute the likelihood for a given value of the parameters, we first propagate the neutrino fluxes from the atmosphere to the detector for both neutrinos and antineutrinos, then we weigh the events from the IceCube Monte Carlo with the propagated flux, which is a function of the true neutrino energy $E_\nu$ and the zenith angle $\theta_z$, and we construct two-dimensional histograms as a function of the reconstructed variables: $E^{\rm rec}_\mu$ and $\theta^{\rm rec}_z$ (using 10 bins in muon energy and 60 angular bins).
 
All the credible intervals we indicate correspond to the highest posterior density interval (i.e., the shortest interval on a posterior probability density for a given confidence level) for one-dimensional marginalized distributions. As a consequence, these intervals always include the point with the highest posterior density, which we also indicate, as a reference, for each quantity. Unless otherwise stated, the credible intervals are all provided for an integrated 68\% posterior probability.

%%%%%%%%%%%%
\begin{figure*}
%	\begin{center}
		\begin{tabular}{cc}
			\includegraphics[width=0.5\textwidth]{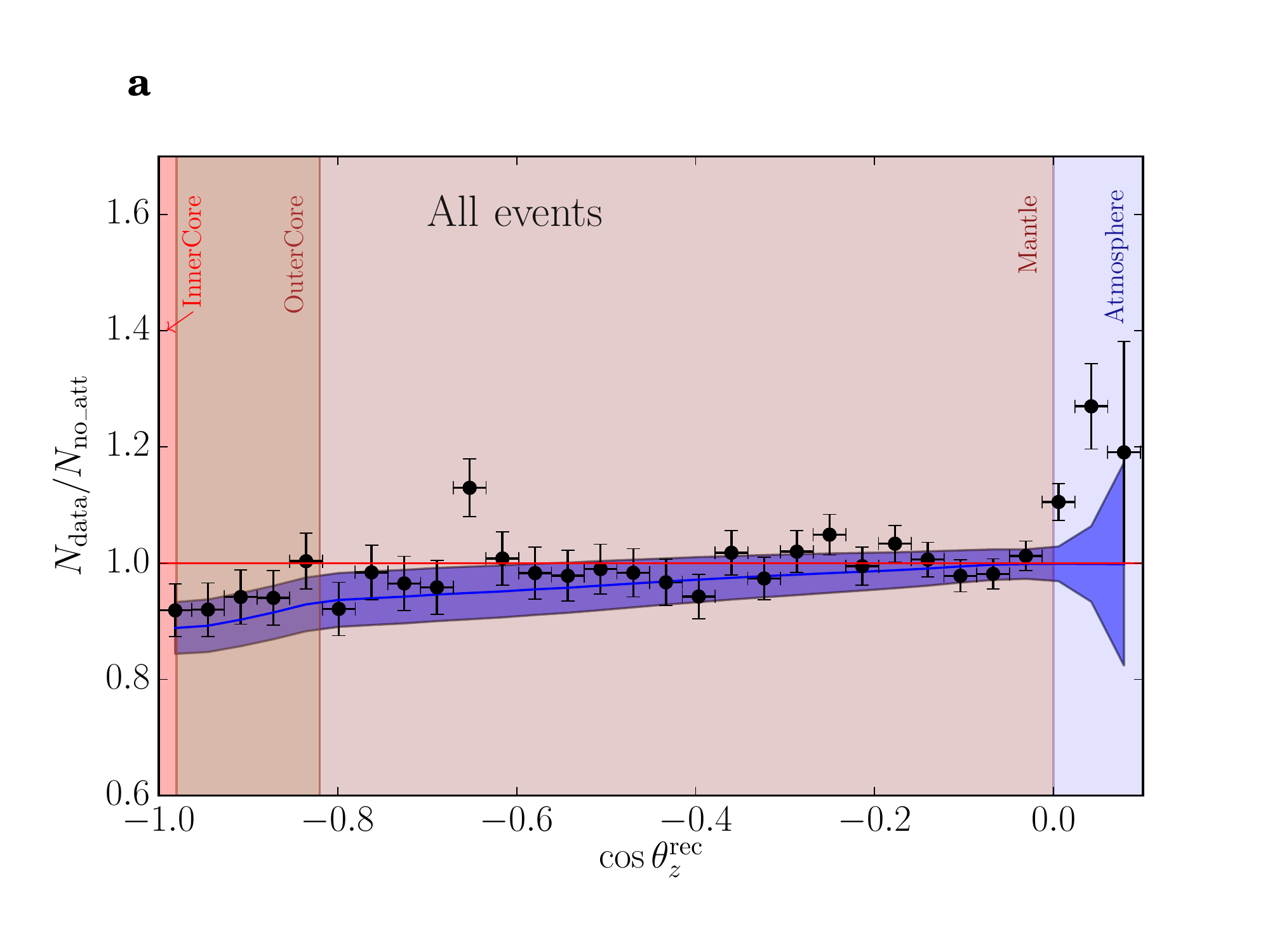} &
			\includegraphics[width=0.5\textwidth]{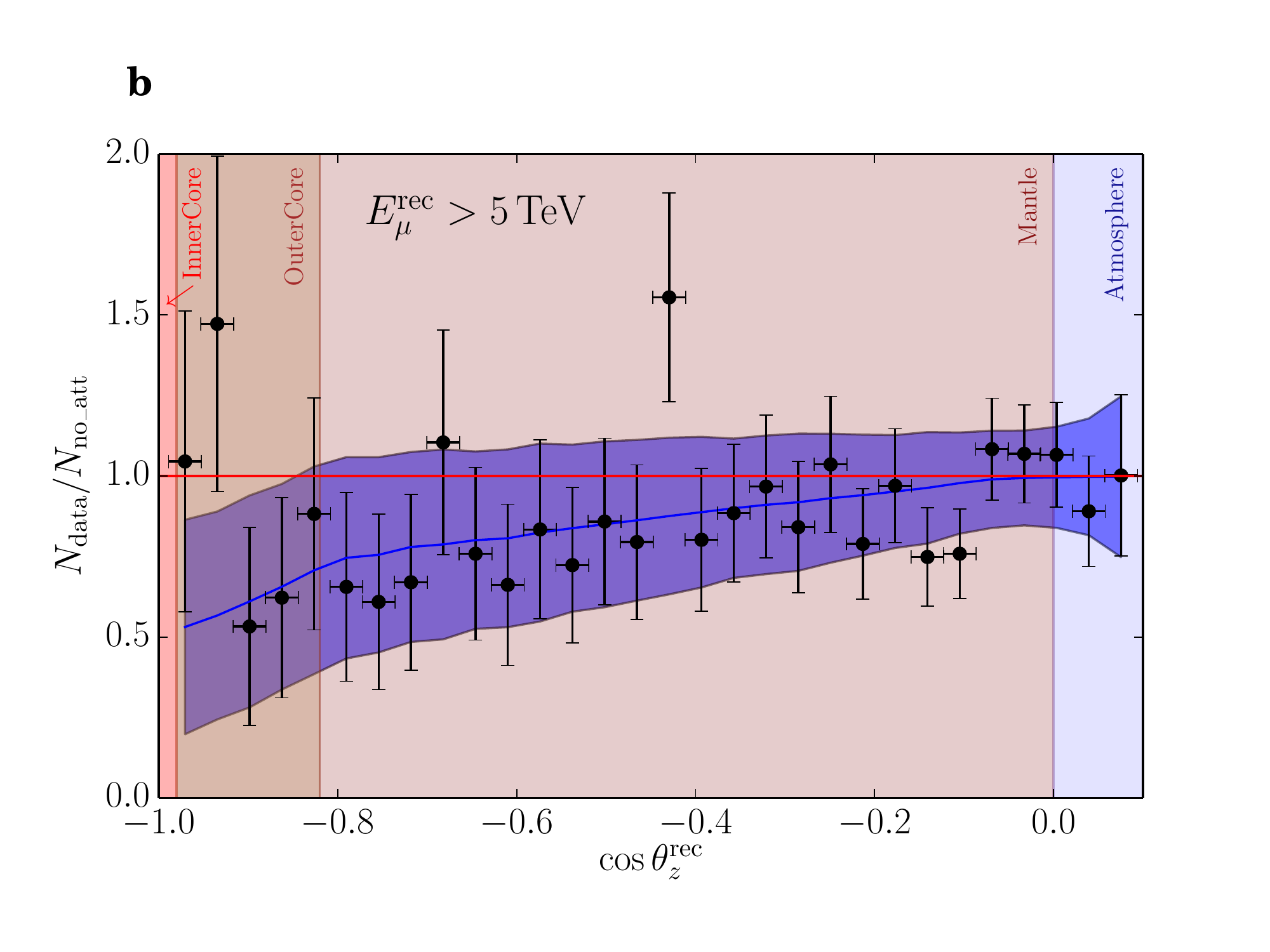}
		\end{tabular}
		\caption{
		{\bf Ratio of the number of observed events in the IC86 sample to the number of expected events without including Earth attenuation.}
			{\bf a,} Zenith distribution of the ratio, including all events in the IC86 sample. {\bf b,} Zenith distribution of the ratio, but only considering events with a minimum reconstructed muon energy of 5~TeV. In both panels, the solid blue line represents the expectation using the PREM~\cite{Dziewonski:1981xy} for the density profile, with its statistical expected error represented by the blue band. 
		}
		\label{fig:zenithratio}
%	\end{center}
\end{figure*}
%%%%%%%%%%%%

\subsection{Energy and zenith distribution}

The energy and zenith distributions of the IC86 sample are shown in Fig.~\ref{fig:zenith}b. Since the atmospheric neutrino spectrum is a steeply falling function of the energy and, for the lowest energies, the neutrino absorption length is much larger than the Earth's diameter, most of the neutrinos in the sample do not undergo significant absorption. Therefore, the distribution of the full sample is very similar to the atmospheric neutrino distribution at the Earth's surface, which is more peaked towards the horizon~\cite{Gaisser:2002jj}.  For higher energies, however, the observed event spectrum corresponding to up-going neutrinos with the longest trajectories through the Earth ($\cos \theta_z^{\rm rec} \sim -1$) is suppressed with respect to the zenith-symmetric flux corresponding to down-going neutrinos that only propagate a few tens of kilometers without crossing the Earth ($\cos \theta_z^{\rm rec} \sim 1$). The effect is more pronounced for neutrinos with higher energies and for those with longer propagation paths in the Earth, as they have a larger probability of interaction. Hence, by studying the zenith and energy distributions of the atmospheric neutrino flux and by comparing them with the flux without attenuation, information on the Earth's density profile can be extracted. All events are useful, though: the events with the lowest energies or more horizontal trajectories serve us to fix the overall normalization and zenith distribution of the unattenuated atmospheric neutrino flux. 

To illustrate how to remove the intrinsic zenith dependence on the atmospheric neutrino flux when comparing with the observed data, we depict the ratio of the observed number of events to the expected one in the case of no absorption, $N_{\rm data}/N_{\rm no\_att}$, as a function of the zenith angle. If all energies in the IC86 sample are considered (Fig.~\ref{fig:zenithratio}a), statistics is dominated by the low-energy events and the maximum observed suppression is at the 10\% level or below. For events with energies above 5~TeV (Fig.~\ref{fig:zenithratio}b), however, the suppression in some of the most vertical angular bins ($\cos \theta_z^{\rm rec} < -0.6$) is up to 50\%. For all energies, the suppression is larger for more vertical trajectories, which imply a longer propagation path. As an indication, we also show the expectations for the central value and the 1$\sigma$ statistical error of this ratio using the one-dimensional Preliminary Reference Earth Model (PREM)~\cite{Dziewonski:1981xy}.

\section{Results}

\subsection{Using one-year IC86}

\begin{figure}[t]
%	\begin{center}
		\includegraphics[width=0.8\textwidth]{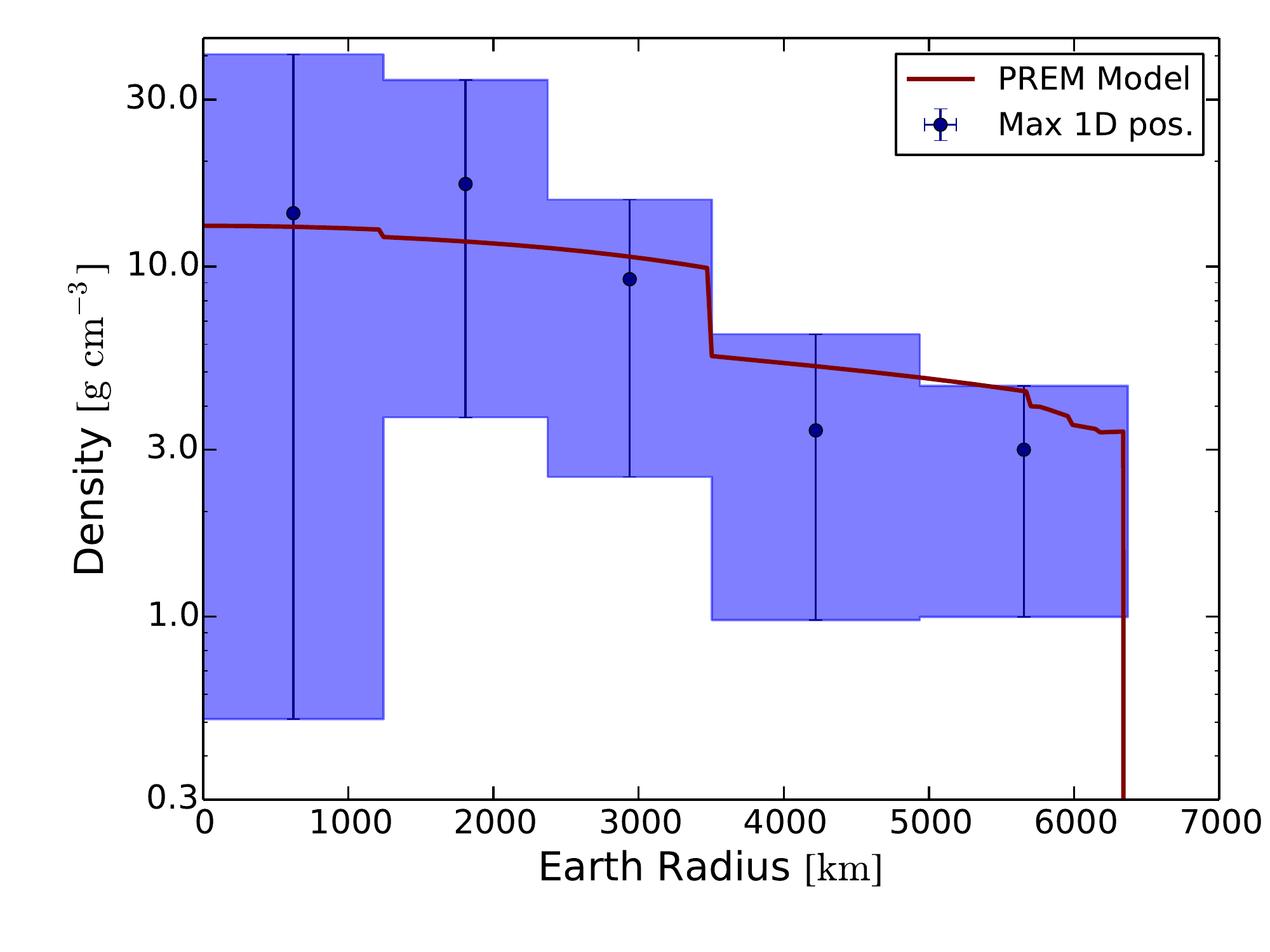} 
		\caption{ {\bf Fit of the density profile of the Earth with IC86 data.}
			Error bars represent 68\% credible intervals (highest one-dimensional marginalized posterior density intervals) and the points with the highest one-dimensional marginalized posterior density are indicated by dots. We assume the Earth is divided into five concentric layers of constant density. The purple curve represents the PREM density profile.
		}
		\label{fig:fitprofile}
%	\end{center}
\end{figure}

We have parametrized the Earth's density with a one-dimensional five-layer profile with constant density in each of the layers (Fig.~\ref{fig:zenith}a). One of the edges is chosen at the core-mantle boundary and another one at the inner core-outer core boundary, so that we select three layers in the core (one for the inner core and two for the outer core) and two layers in the mantle. We have checked that, with this number of layers, current data are not yet sensitive to the particular profile within a given layer (see Figs.~\ref{fig:otherprof} and~\ref{fig:triangleplototherprof} and columns 6-7 of Tab.~\ref{tab:results}) and, therefore, there is no expected gain when using more layers or a more realistic density profile. We fit the {\it average} density of each of the layers, which is allowed to vary freely, and obtain our main result, the first one-dimensional Earth's density profile measured by means of weak interactions (Fig.~\ref{fig:fitprofile}). With one-year statistics the uncertainties are large but, yet, compatible with results from geophysical methods within 68\% credible interval. Notice that these results are obtained from one-dimensional marginalized posterior probability distributions and correlations among all the parameters in the fit (five densities and four nuisance parameters) are not shown here. They give, therefore, a conservative representation of allowed ranges for the density of individual layers.

%%%%%%%%%%%%
%%%%%%%%%%%%

%%%%%%%%%%%%
\begin{figure*}[t]
%	\begin{center}
		\includegraphics[width=0.74\textwidth]{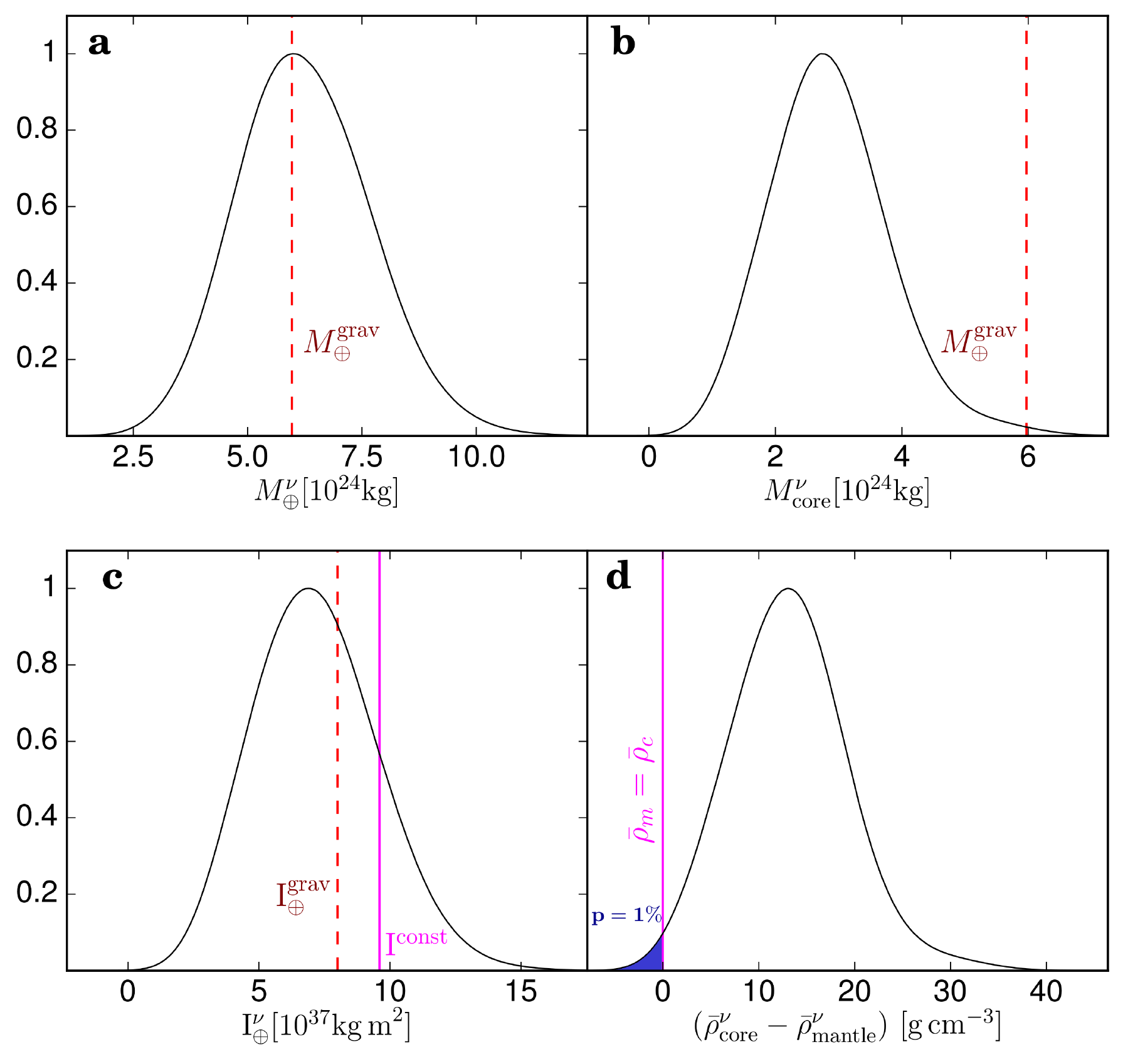} 
		\caption{
		 {\bf Earth measurements from neutrino tomography.}
			{\bf a,} Posterior probability for the Earth's mass (black solid curve) compared to its gravitational measurement, $M_\oplus^{\rm grav}$ (red dashed line). 
			{\bf b,} Posterior probability for the mass of the Earth's core (black solid curve) compared to the gravitational measurement of the Earth's mass, $M_\oplus^{\rm grav}$ (red dashed line). 
			{\bf c,} Posterior probability for the Earth's mean moment of inertia (black solid curve) compared to its gravitational measurement, $I_\oplus^{\rm grav}$ (red dashed line). The value for the moment of inertia corresponding to a homogeneous Earth ($0.4 \, M_\oplus^{\rm grav} \, R_\oplus^2$), assuming the gravitational mass determination, is also shown (thin magenta solid line).
			{\bf d,} Posterior probability for the difference $\bar\rho_{\rm core}^\nu - \bar\rho_{\rm mantle}^\nu$ between the average core density, $\bar \rho_{\rm core}^\nu$, and the average mantle density, $\bar \rho_{\rm mantle}^\nu$. We also indicate the point where $\bar\rho_{\rm core}^\nu = \bar\rho_{\rm mantle}^\nu$ (thin magenta solid line) and the $p-$value for a denser mantle (blue region). 
		}
		\label{fig:results}
%	\end{center}
\end{figure*}
%%%%%%%%%%%%

From the results of the fit, we compute the mass of the Earth as {\it weighted} by neutrinos and obtain $M_\oplus^{\nu} = \left( 6.0 ^{+1.6}_{-1.3} \right) \times 10^{24}$~kg (Fig.~\ref{fig:results}a), to be compared to the most precise gravitational measurement up to date, $M_\oplus^{\rm grav} = \left( 5.9722 \pm 0.0006 \right) \times 10^{24}$~kg~\cite{Luzum:2011, AAlmanac}. Clearly, albeit within large uncertainties, both results are in very good agreement. 

We can also estimate the mass of the Earth's core, a parameter that may be useful (as soon as statistical errors will decrease) as an input for geophysical measurements of the Earth's density profile. The result for this quantity is $M^{\nu}_{\rm core} = \left( 2.72 ^{+0.97}_{-0.89} \right) \times 10^{24}$~kg, which is slightly larger than the result from geophysical density models, that estimate the mass of the core to be $\sim 33\%$ of the total mass of the Earth (see Fig.~\ref{fig:results}b).

From our measurement of the one-dimensional density profile we can determine the Earth's moment of inertia, for which we get $I_\oplus^{\nu} = \left( 6.9 \pm 2.4 \right) \times 10^{37}$~kg~m$^2$~(Fig.~\ref{fig:results}c), in agreement with the current (gravitationally inferred) measurement of the mean moment of inertia, $I_\oplus^{\rm grav} = \left( 8.01736 \pm 0.00097 \right) \times 10^{37}$~kg~m$^2$~\cite{Chen:2015}. The smaller moment of inertia from neutrino data, as compared to gravitational measurements, implies a central value with a larger departure from homogeneity, as shown in Fig.~\ref{fig:results}c (even though they are fully compatible between each other due to the yet large uncertainties). 

Another piece of information regarding the Earth's interior that we can extract from the currently available data is to confirm the core is denser than the mantle, which is necessary for the Earth to be gravitationally stable. Notice that, implicitly, this is a strong assessment in favor of a non-homogeneous Earth (something that was expected to be possible to proof at $3 \sigma$ after ten years of IceCube data~\cite{GonzalezGarcia:2007gg} and seems to be already established at more than $2 \sigma$ just using IC86 alone). We determine the difference between the average density within the two layers we divide the mantle into,  $\bar \rho_{\rm mantle}$, and the average density within the three layers corresponding to the core, $\bar \rho_{\rm core}$. The result for this difference, measured by weak interactions, is $\left(\bar\rho_{\rm core}^\nu - \bar\rho_{\rm mantle}^\nu \right) = 13.1 ^{+5.8}_{-6.3}$~g/cm$^3$ (Fig.~\ref{fig:results}d). From this result, a denser Earth's mantle has a $p-$value of 0.011 for our default model of the atmospheric neutrino flux. 

\begin{figure*}[t]
	\begin{center}
		\includegraphics[width=\textwidth]{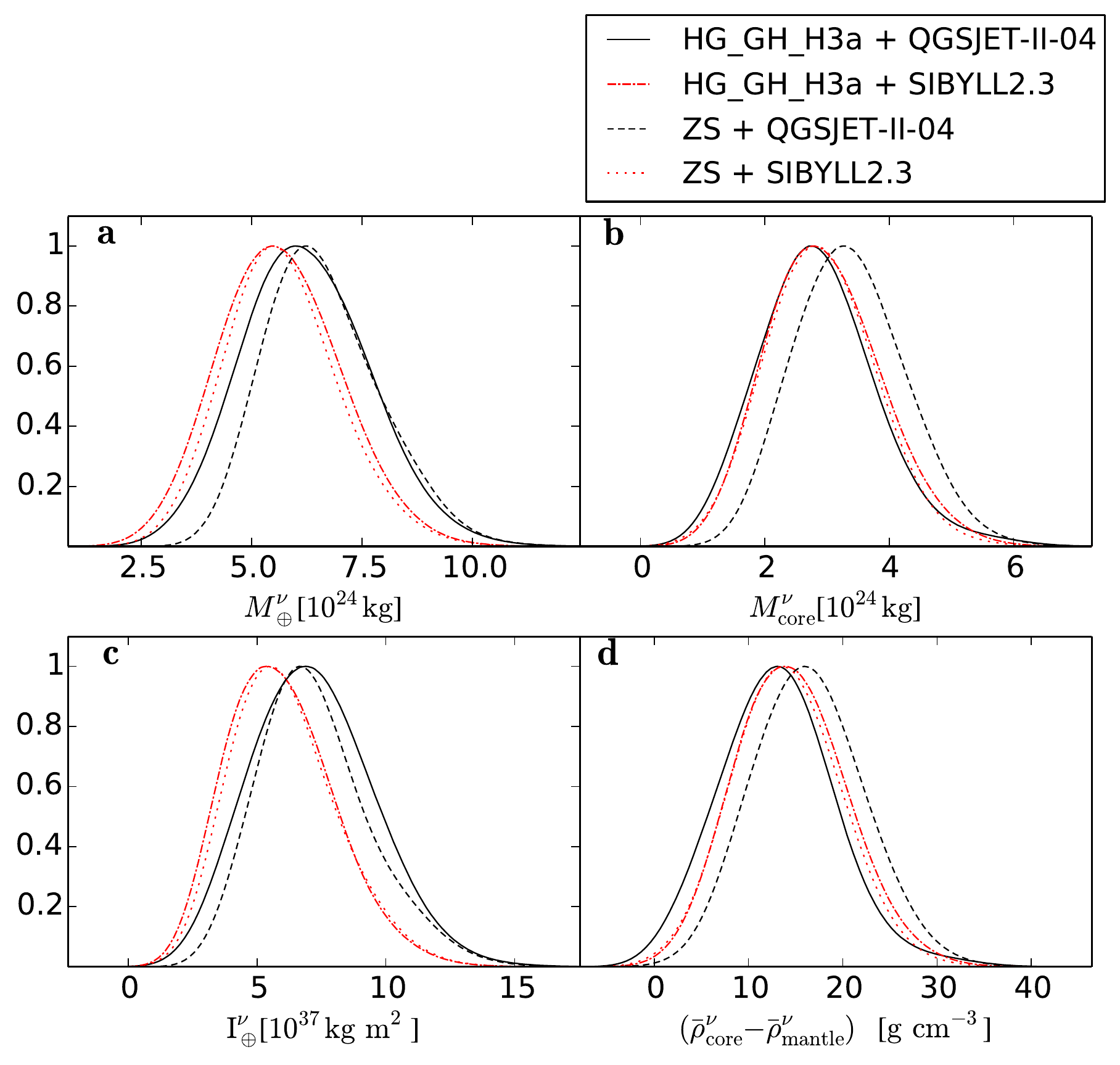} 
		\caption{ {\bf Systematic uncertainties among different atmospheric neutrino fluxes.} Posterior probability distributions (normalized such that the maximum is 1) of the measured quantities for the Earth using neutrino tomography for four different atmospheric neutrino fluxes, resulting from the combinations of two primary cosmic-ray fluxes: the combined Honda-Gaisser model and Gaisser-Hillas H3a correction (HG-GH-H3a) and the Zatsepin-Sokolskaya (ZS) spectrum, and two hadronic models, QGSJET-II-4 and SIBYILL2.3. All measurements are dominated by statistical uncertainties, being the systematics introduced by differences among atmospheric neutrino fluxes a subdominant effect. {\bf a,} Earth's mass. {\bf b,} Earth's core mass. {\bf c,} Earth's moment of inertia. {\bf d,} Difference of the average density between the Earth's core and mantle. The $p-$value for a mantle denser than the core corresponds to the area in the region where $\bar\rho_{\rm core}^\nu \leq \bar\rho_{\rm mantle}^\nu$. Our default model, HG-HG-H3a + QGSJET-II-4, has the larger $p-$value.
		}
		\label{fig:otheratm}
	\end{center}
\end{figure*}

\begin{figure*}
	\includegraphics[width=\textwidth]{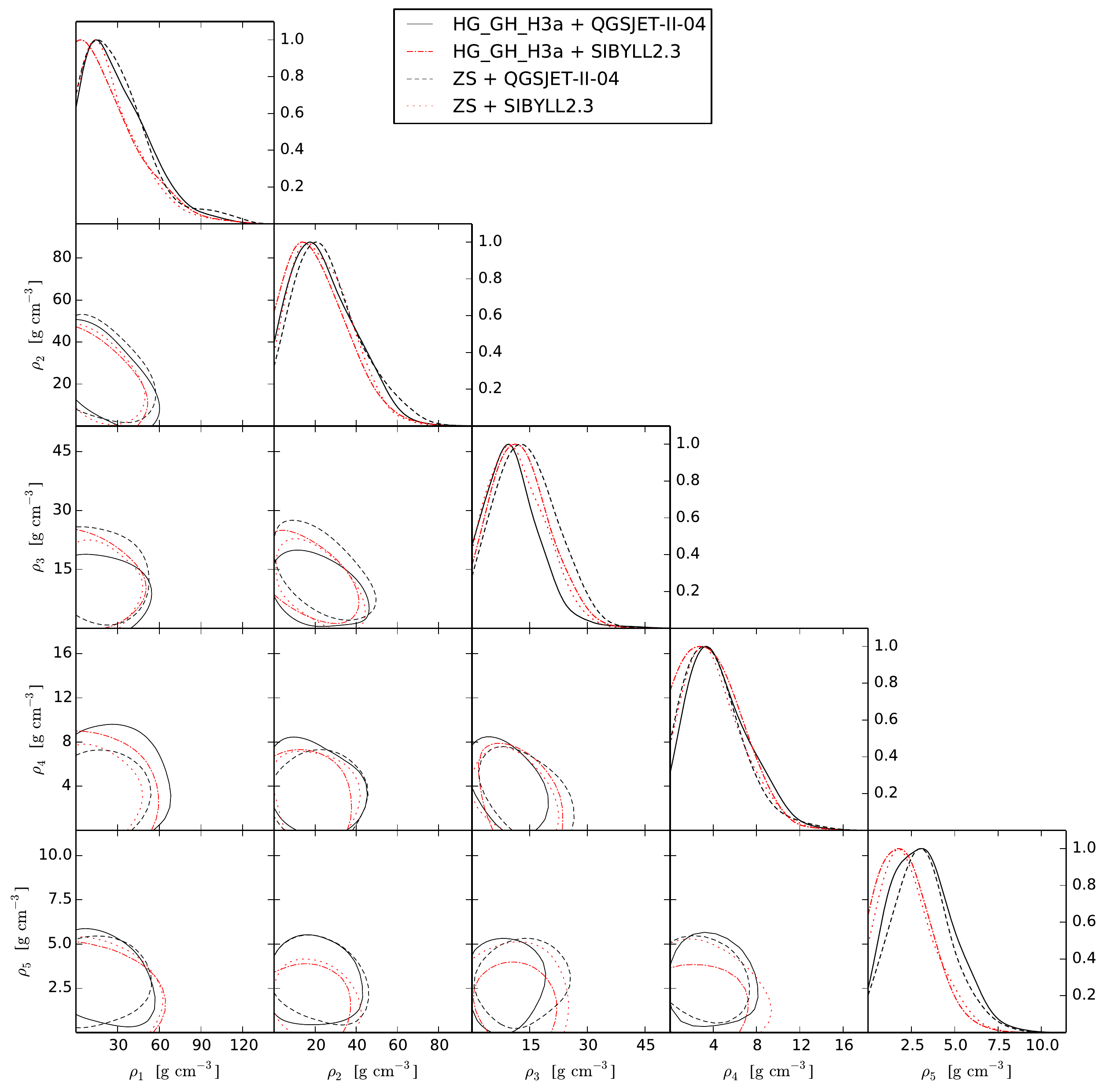} 
	\caption{ {\bf Posterior 68\% probability contours for the densities of the Earth's layers.} We model the Earth with a piecewise flat profile, where each of the layers is described with constant density: $\rho_1$ corresponds to the inner core, $\rho_2$ and $\rho_3$ to the equal-thickness layers of the outer core, $\rho_4$ and $\rho_5$ to the equal-thickness layers of the mantle. We show the results for the four different combinations of cosmic-ray spectrum and hadronic models in  Fig.~\ref{fig:otheratm}. With current data, the results are dominated by statistical uncertainties. On the rightmost panels, we depict the one-dimensional marginalized posterior probability distribution of the density of the layer corresponding to each column, normalized such that the maximum is 1.
	}
	\label{fig:triangleplototheratm}
\end{figure*}

\begin{figure*}[t]
	\begin{center}
		\includegraphics[width=\textwidth]{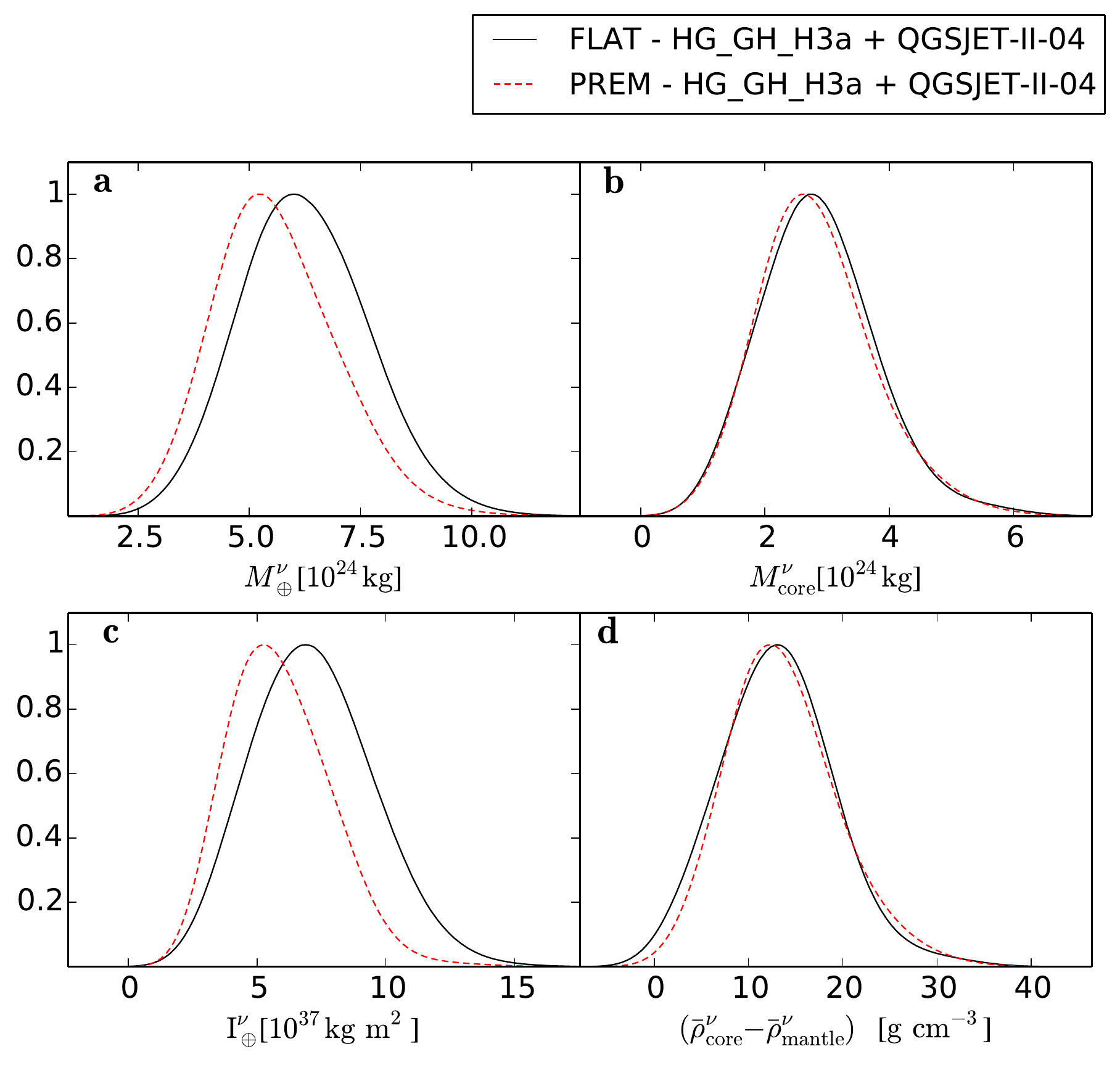} 
		\caption{ {\bf Systematic uncertainties between Earth density profiles.}  Posterior probability distributions (normalized such that the maximum is 1) of the measured quantities for the Earth using neutrino tomography for two different Earth's density profiles: a piecewise profile with five layers of constant density (as in Fig.~\ref{fig:otheratm}) and a five-layer model following the PREM profile. In all cases we use our default atmospheric neutrino fluxes: the combination of the Honda-Gaisser model with the Gaisser-Hillas H3a correction (HG-GH-H3a) and the QGSJET-II-04 hadronic model. {\bf a,} Earth's mass. {\bf b,} Earth's core mass. {\bf c,} Earth's moment of inertia. {\bf d,} Difference of the average density between the Earth's core and mantle.
		}
		\label{fig:otherprof}
	\end{center}
\end{figure*}

\begin{figure*}
	\includegraphics[width=\textwidth]{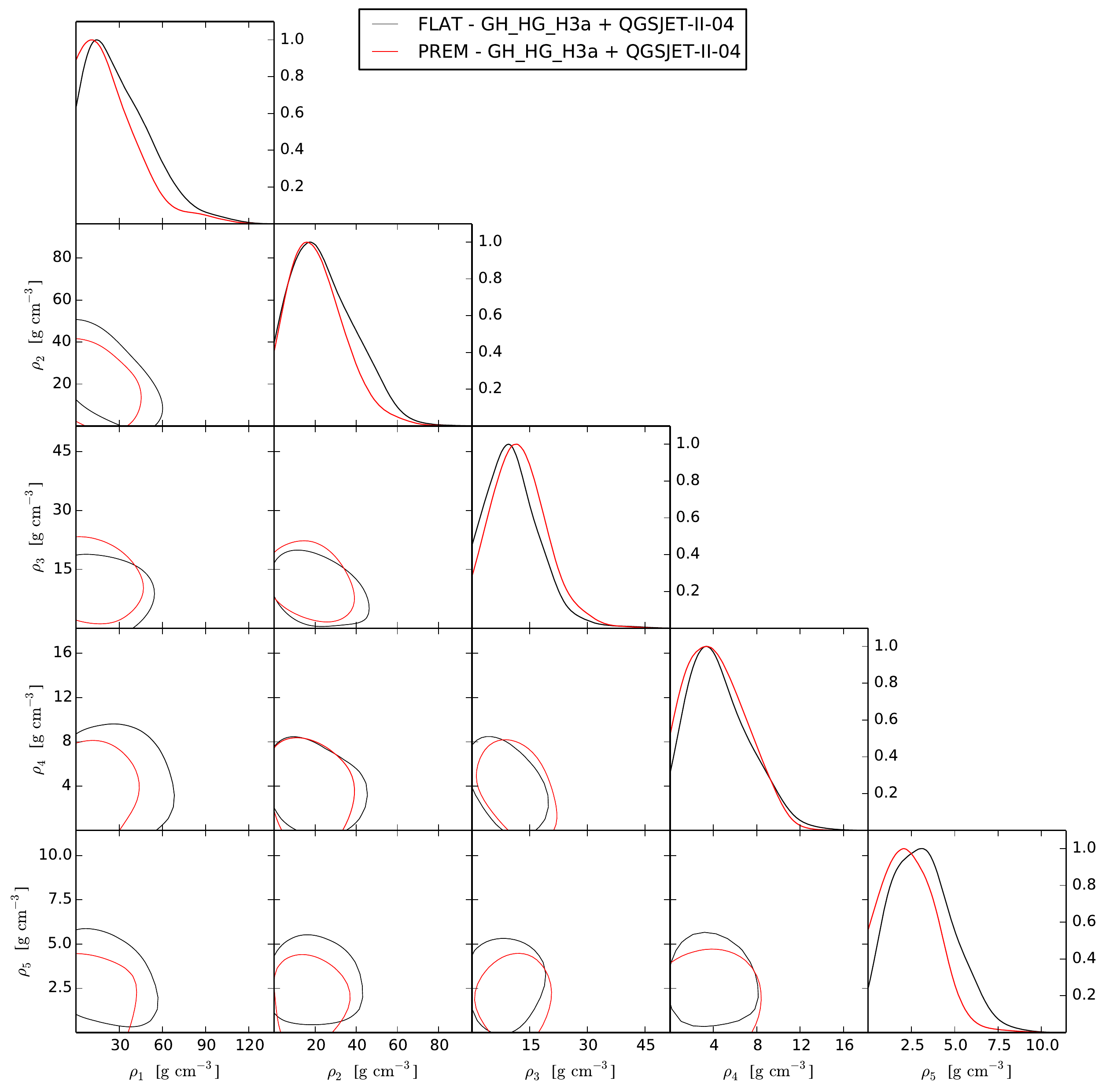} 
	\caption{ {\bf Posterior 68\% probability contours for the densities of the five layers.} We show the results for the densities of the layers corresponding to two different density profiles: a piecewise profile with five layers of constant density (as in Fig.~\ref{fig:triangleplototheratm}) and a five-layer model following the PREM profile. For the latter (non-constant density within the layers), the densities shown correspond to the value at the center of each layer. For the atmospheric neutrino fluxes, we consider the combination of the Honda-Gaisser model with the Gaisser-Hillas H3a correction (HG-GH-H3a) and the QGSJET-II-04 hadronic model. With current data, the results are dominated by statistical uncertainties. On the rightmost panels, we depict the one-dimensional marginalized posterior probability distribution of the parameter corresponding to each column, normalized such that the maximum is 1.
	}
	\label{fig:triangleplototherprof}
\end{figure*}

As a test of consistency and as a matter of accounting for further systematic uncertainties, all observables have also been computed for other atmospheric neutrino fluxes. The results for all these cases are shown in Figs.~\ref{fig:otheratm} and~\ref{fig:triangleplototheratm} and in columns 2-5 of Tab.~\ref{tab:results}. This overall systematic uncertainty results into shifts of the allowed range for the fitted and derived quantities of about $\sim (20-30)\%$. In addition, we have also performed an analysis using different modeling of the inner structure of the Earth. The results for the different cases is shown in Figs.~\ref{fig:otherprof} and~\ref{fig:triangleplototherprof} and in columns 6-7 of Tab.~\ref{tab:results}. Nevertheless, to understand the importance of these external constraints, we have performed an analysis of the present statistical sample including the total mass of the Earth and its moment of inertia as external priors. We have found that this procedure constrains the mantle density (mainly the outermost mantle layer) with a better precision than what can be done with one year of high-energy neutrino data, whereas these priors have a rather small impact on our results for the inner and outer core densities, given the already large uncertainties. 

We have also estimated the contribution of these nuisance parameters to the error budget of the four derived quantities presented in Tab.~\ref{tab:results} (the Earth's mass, the Earth's core mass, the Earth's moment of inertia and the difference in average density of the core and mantle)  by comparing our results with the outcome of a fit where the four nuisance parameters have been fixed to their corresponding best-fit values. We have found that these systematic errors contribute to approximately 30\% of the error on the derived quantities, $M^\nu_\oplus$, $M^\nu_{\rm core}$, $I^\nu_\oplus$ and $\bar \rho^\nu_{\rm core} - \bar \rho^\nu_{\rm mantle}$.

\begin{table*}[t]
	\def\arraystretch{3}
	\centering
	\resizebox{\textwidth}{!}{\begin{tabular}{| c || c | c | c | c | c |}
			\hline
			\multicolumn{1}{|c||}{} & \multicolumn{4}{|c|}{\Large Piecewise flat Earth's profile} & \multicolumn{1}{c|}{\Large PREM Earth's profile} \\ \hline
			& \bf HG-GH-H3a + QGSJET-II-04 & HG-GH-H3a + SIBYLL2.3 & ZS + QGSJET-II-04 & ZS + SIBYLL2.3 & HG-GH-H3a + QGSJET-II-04 \\ \hline
			$M_\oplus^\nu$ $[10^{24}~{\rm kg}]$ & \Large $6.0 ^{+1.6}_{-1.3}$ & \Large $5.5 ^{+1.5}_{-1.3}$ & \Large $6.2 ^{+1.4}_{-1.2}$ & \Large $5.5 ^{+1.3}_{-1.2}$ & \Large $5.3 ^{+1.5}_{-1.3}$ \\ \hline
			$M_{\rm core}^\nu$ $[10^{24}~{\rm kg}]$& \Large $2.72 ^{+0.97}_{-0.89}$ & \Large $2.79 ^{+0.98}_{-0.85}$ & \Large $3.27 ^{+0.92}_{-0.89}$ & \Large $2.84 ^{+0.89}_{-0.88}$ & \Large $2.62 ^{+0.97}_{-0.84}$ \\ \hline
			$I_\oplus^\nu$ $[10^{37}~{\rm kg}~{\rm cm}^2]$& \Large $6.9 \pm 2.4$ & \Large $5.4 ^{+2.3}_{-1.9}$ & \Large $6.7 ^{+2.3}_{-2.0}$ & \Large $5.5 ^{+2.2}_{-1.9}$ & \Large $5.3 ^{+2.3}_{-1.7}$ \\ \hline
			$\bar\rho_{\rm core}^\nu - \bar\rho_{\rm mantle}^\nu$ $[{\rm g/}{\rm cm}^3]$ & \Large $13.1 ^{+5.8}_{-6.3}$ & \Large $14.0 ^{+6.0}_{-5.9}$ & \Large $15.9 ^{+6.0}_{-5.9}$ & \Large $13.5^{+6.1}_{-5.5}$ & \Large $12.3 ^{+6.3}_{-5.4}$ \\ \hline
			$p-{\rm value}$ &  \multirow{2}{*}{\Large $1.1 \times 10^{-2}$} &  \multirow{2}{*}{\Large $2.4 \times 10^{-3}$} &  \multirow{2}{*}{\Large $9.4 \times 10^{-4}$} &  \multirow{2}{*}{\Large $4.6 \times 10^{-3}$} &  \multirow{2}{*}{\Large $3.8 \times 10^{-3}$} \\
			mantle denser than core & & & & & \\ \hline
	\end{tabular}}
	\caption{ {\bf Results from neutrino tomography using one year of data (IC86 sample).} Here we indicate the maximum of the posterior probability and the 68\% credible interval (defined as the highest one-dimensional marginalized posterior density interval) for each derived quantity: the Earth's mass, the Earth's core mass, the Earth's moment of inertia, and the difference in average density of the core and mantle. We also indicate the $p-$value for a mantle denser than the core ($\bar\rho_{\rm core}^\nu \leq \bar\rho_{\rm  mantle}^\nu$). We show the results for four atmospheric neutrino fluxes assuming a piecewise profile with five constant-density layers and for a PREM-like profile with five layers, and the combination of the Honda-Gaisser model with the Gaisser-Hillas H3a correction (HG-GH-H3a) and the QGSJET-II-04 hadronic model.
	}
	\label{tab:results}
\end{table*}

%%%%%%%%%%%%
%%%%%%%%%%%%

%%%%%%%%%%%%
%%%%%%%%%%%%

\begin{figure*}
	\includegraphics[width=\textwidth]{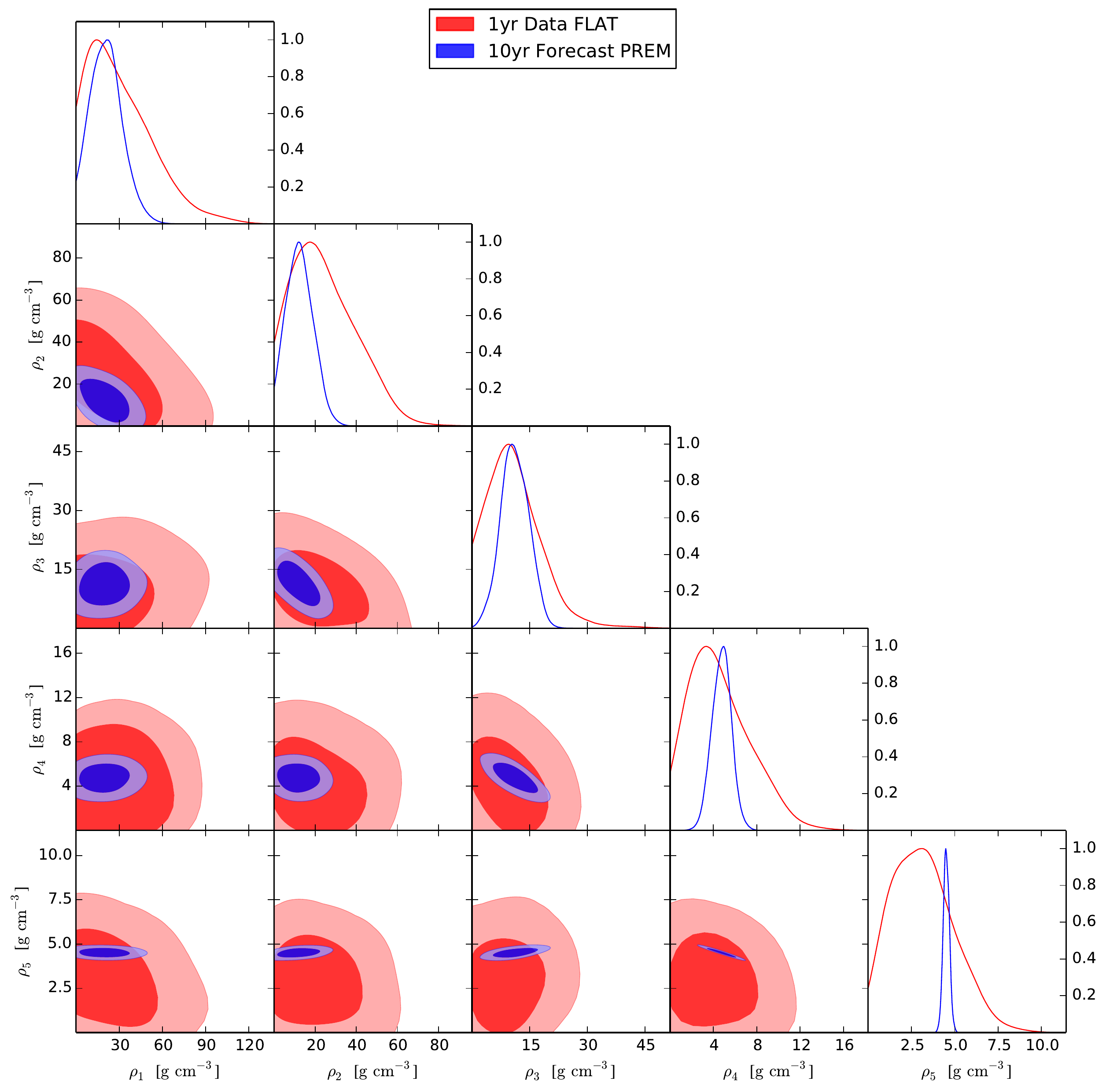} 
	\caption{{\bf Ten-year forecast versus current results.} Posterior 68\% and 95\% probability contours for the densities of the five constant-density layers: $\rho_1$ corresponds to the inner core, $\rho_2$ and $\rho_3$ to the equal-thickness layers of the outer core, $\rho_4$ and $\rho_5$ to the equal-thickness layers of the mantle. We compare the results obtained with the current one-year IC86 data assuming a piecewise flat profile (red contours), with the forecast for 10 years (blue contours). For the forecast analysis, we simulate the future data assuming the PREM density profile and fit it with a model with five layers following the PREM profile in each layer (but with free normalization), so that the values indicated in the plots correspond to the central value in each of the layers. In all cases, for the atmospheric neutrino fluxes, we consider the combination of the Honda-Gaisser model with the Gaisser-Hillas H3a correction (HG-GH-H3a) and the QGSJET-II-04 hadronic model. For the forecast, we use the same systematic uncertainties that we have used throughout the paper. However, it is reasonable to think that they would be improved in the future. The outcome of the forecast is that, whereas with current data the results are dominated by statistical uncertainties, impressive improvements can be achieved already with a factor of ten larger statistics. The mantle density would be known with a much better precision, and our understanding of the Earth's core will increase significantly. Finally, note that currently more than seven years of data have already been collected, although data are not publicly available in the adequate form to perform this kind of analysis. On the rightmost panels, we depict the one-dimensional marginalized posterior probability distribution of the density of the layer corresponding to each column, normalized such that the maximum is 1.
	}
	\label{fig:triangleplotforecast}
\end{figure*}

\begin{figure*}
	\includegraphics[width=0.8\textwidth]{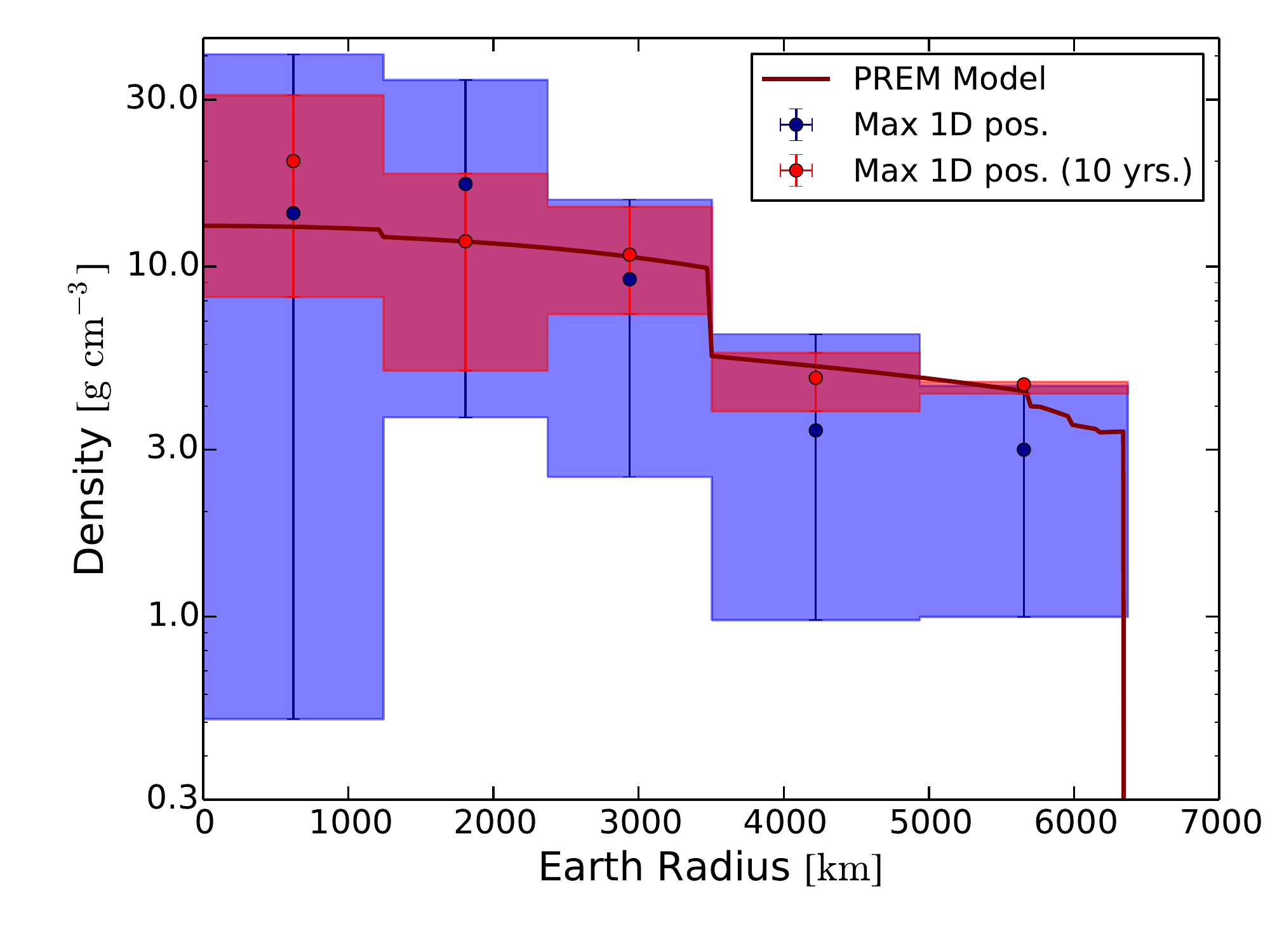} 
	\caption{ {\bf Ten-year forecast versus current results: density profile.} Fitted one-dimensional Earth's density profile with error bars representing 68\% credible intervals (defined as the highest one-dimensional marginalized posterior density intervals) and with the points with the highest one-dimensional marginalized posterior density indicated by dots. The blue bands and points represent the results obtained using current one-year (IC86) data and assuming the Earth is divided into five concentric layers of constant density (same as Fig.~3). The red bands and points represent the expected results after ten years of observation. We have simulated the future data assuming the PREM density profile and fitted it with a model with five layers following the PREM profile in each layer (but with free normalization), so that the values indicated in the plots correspond to the central value in each of the layers. For the atmospheric neutrino fluxes, we consider the combination of the Honda-Gaisser model with the Gaisser-Hillas H3a correction (HG-GH-H3a) and the QGSJET-II-04 hadronic model. The purple curve represents the PREM density profile. Note that these results are obtained from one-dimensional marginalized posterior probability distributions and, therefore, correlations among all the parameters in the fit (five densities and four nuisance parameters) cannot be represented here. They give, therefore, a conservative representation of the allowed ranges for the density of individual layers.
	}
	\label{fig:fitprofileforecast}
\end{figure*}

\subsection{Ten-year forecast}

It is interesting to get an idea of how the measurements shown in this paper may improve as soon as more data will become available. For this reason, we also compare our results with the outcome of an analysis performed assuming ten years of data. For this forecast analysis, we consider the combination of the Honda-Gaisser model with the Gaisser-Hillas H3a correction (HG-GH-H3a) and the QGSJET-II-04 hadronic model for the atmospheric neutrino flux and we simulate the future data assuming the PREM density profile. Simulated data are subsequently fitted (using the same atmospheric neutrino flux) with a five-layer model, as in the current analysis using the IC86 sample, albeit with a density following the PREM profile within each layer. This approach is used to avoid coarse binning with higher statistics, in the case of a piecewise flat profile. Although with current data, considering five layers with constant density is equally good as assuming a more realistic profile as the PREM model (see Figs.~\ref{fig:otherprof} and~\ref{fig:triangleplototherprof} and Tab.~\ref{tab:results}), with more data, a finer modeling of the Earth with more than five layers or a more accurate profile within layers would be certainly needed. The results of the comparison are shown in Figs.~\ref{fig:triangleplotforecast} and~\ref{fig:fitprofileforecast}.

In our default forecast analysis, as described above, we assume that future data will come along with a better determination of the atmospheric neutrino flux model and that, therefore, a fit of the forecast data can be performed using only the flux model used to generate the data themselves. Nevertheless, we have also studied the impact of the discrete choices of primary flux and hadronic model on the forecast. We have used different combinations to generate and fit data. We have found that, for some combinations of fluxes, the results of the fit give a statistically significant disagreement with the gravitational measurement of the Earth's mass and of the Earth's moment of inertia (whereas the Earth's core mass and the core-mantle density jump are little affected by the choice of flux model). However, given that Earth absorption only takes place for the highest energy neutrinos and the flux is a steeply falling power-law spectrum, the part of the sample with the largest statistics, that actually allows us to improve our knowledge of the atmospheric neutrino flux, comes from energies very little affected by the passage of neutrinos through the Earth. Therefore, it is not clear that the addition of external constraints on the Earth's mass and moment of inertia, in the analysis of future data, could falsify some of the choices of neutrino flux models. However, we recall that uncertainties in the flux models will also be reduced by other complementary future measurements (such as, for example, the measurement of the atmospheric muon flux, improved cosmic-ray measurements, better understanding of hadronic interactions, measurements of atmospheric neutrino fluxes at lower energies and even at similar energies for down-going neutrinos of different flavors). For the forecast to properly take into account potential improvements on the ingredients of the analysis, other different type of data would certainly have to be included, going beyond the scope of this letter.

We have also performed more detailed ten-year forecast analyses, considering different density profiles within each layer (either flat or following the PREM) and several configurations of layers. From these analyses we have verified that: (1) the statistical error in the outer mantle layers could go down to a few percent; (2) the statistical error in the inner mantle layers will get reduced down to around 10\%; (3) a finer description (more layers) of the one-dimensional Earth's profile than the one used in the present work will be needed. It is not yet clear if with ten years of data it will be possible to determine the location of the core-mantle boundary just by looking at high-energy neutrino data, but what is clear from the forecast analysis is that a simple five-layer Earth's model would not be the optimal one to analyze the data and more layers would represent a better description of the density profile. For example, we have checked that the results of a five-layer fit would be affected by the choice of the profile within layers, as for instance, flat layers (constant density within each layer) versus layers with a density profile that follows the PREM.

\section{Conclusions}

At high enough energies (few TeV), the passing of neutrinos through the Earth is sensitive to the number density of nucleons and, therefore, this test represents an effective counting of nucleons in the Earth. Unlike gravitational methods, the estimation of the Earth's mass with neutrinos relies purely on weak interactions and on the values of the nucleon masses. Conceptually, this is a completely different method from gravitational ones. We have shown that, using the publicly available data from the IceCube neutrino telescope, this method starts being feasible. Future data will significantly improve the measurements presented here (we remind that more data already collected by IceCube in the same energy range are not yet publicly available in the format required to perform this analysis, but hopefully will be released soon). For this reason, we have also estimated the projected sensitivity with future data (see Figs.~\ref{fig:triangleplotforecast} and~\ref{fig:fitprofileforecast}). 

As a final comment, it is important to stress that a non-gravitational measurement of the Earth mass, as it is the one presented here, could also probe that all the matter that contributes to the Earth gravitational field is baryonic matter (protons, neutrons and electrons). With current neutrino data, however, a small fraction
in the form of (non weakly-interacting) dark matter, which would not attenuate the passage of neutrinos, cannot be yet fully excluded.

\section*{Acknowledgments}
A.D. thanks G.~Cultrera, C.~Piromallo and G.~Soldati for useful discussions. A.D. and J.S. were supported by the Generalitat Valenciana under grant PROMETEO II/2014/050 and by the Spanish MINECO grants FPA2014-57816-P and FPA2017-85985-P. S.P.R. is supported by the Generalitat Valenciana under grant PROMETEOII/2014/049 and by the Spanish MINECO grants FPA2014-54459-P and FPA2017-84543-P. S.P.R. is supported by a Ram\'on y Cajal contract, and also partially by the Portuguese FCT through the CFTP-FCT Unit 777 (PEst-OE/FIS/UI0777/2013). J.S. is also supported by the Spanish MINECO grant FPA2016-76005-C2-1-P, Maria de Maetzu program grant MDM-2014-0367 of ICCUB and research grant 2017-SGR-929. All authors are supported by the European Union's Horizon 2020 research and innovation program under the Marie Sk\l odowska-Curie grant agreements No. 690575 and 674896. \\

\bibliography{NuTomography_biblio}

\end{document}